\documentclass[journal,onecolumn,10pt]{IEEEtran}

\usepackage{caption}
\usepackage{subcaption}
\ifCLASSINFOpdf
  \usepackage[pdftex]{graphicx}
  \graphicspath{{./graph/}}
  \DeclareGraphicsExtensions{.pdf,.jpeg,.png}
\else
  \usepackage[dvips]{graphicx}
  \graphicspath{{./graph/}}
  \DeclareGraphicsExtensions{.eps}
\fi

\ifCLASSINFOpdf
\else
\fi

\usepackage[cmex10]{amsmath,mathtools}
\usepackage{mdwmath}
\usepackage{mdwtab}
\usepackage{amssymb}
\usepackage{amsthm}

\usepackage{algorithm}
\usepackage{algcompatible}

\usepackage{float}

\usepackage{epstopdf}
\usepackage{fancyhdr}
\usepackage{fancyheadings}
\usepackage{multirow}

\usepackage{color}

\begin{document}

\title{LCD: Low Latency Command Dissemination for A Platoon of Vehicles}

\author{\IEEEauthorblockN{Kai Li\IEEEauthorrefmark{1},
Wei Ni\IEEEauthorrefmark{2},
Eduardo~Tovar\IEEEauthorrefmark{1}, and 
Mohsen~Guizani\IEEEauthorrefmark{3}}

\IEEEauthorblockA{\IEEEauthorrefmark{1}Real-Time and Embedded Computing Systems Research Centre (CISTER), Porto, Portugal.\\ 
Email: \{kai\_li, emt\}@isep.ipp.pt.}

\IEEEauthorblockA{\IEEEauthorrefmark{2}Commonwealth Scientific and Industrial Research Organization (CSIRO), Sydney, Australia.\\
Email: wei.ni@csiro.au.}

\IEEEauthorblockA{\IEEEauthorrefmark{3}Department of Electrical and Computer Engineering, University of Idaho, Moscow, USA.\\
Email: mguizani@ieee.org.}}

\maketitle

\IEEEcompsoctitleabstractindextext{%
\begin{abstract}
In a vehicular platoon, a lead vehicle that is responsible for managing the platoon's moving directions and velocity periodically disseminates control commands to following vehicles based on vehicle-to-vehicle communications. However, reducing command dissemination latency with multiple vehicles while ensuring successful message delivery to the tail vehicle is challenging. We propose a new linear dynamic programming algorithm using backward induction and interchange arguments to minimize the dissemination latency of the vehicles. Furthermore, a closed form of dissemination latency in vehicular platoon is obtained by utilizing Markov chain with M/M/1 queuing model. Simulation results confirm that the proposed dynamic programming algorithm improves the dissemination rate by at least 50.9\%, compared to similar algorithms in the literature. Moreover, it also approximates the best performance with the maximum gap of up to 0.2 second in terms of latency. 
\end{abstract}

}

\maketitle

\IEEEdisplaynotcompsoctitleabstractindextext

\IEEEpeerreviewmaketitle

%=============================================================================%
%============================ Section 1 Introduction================================%
\section{Introduction}
\label{sec_intro}
Recent advances in inter-vehicle wireless communications, e.g., Wireless Access in Vehicular Environments (WAVE), or Dedicated Short-Range Communication (DSRC), have enabled a new platoon-based driving paradigm, in which a lead vehicle is driven manually, while the following vehicles follow the lead vehicle in a fully automatic fashion. Every following vehicle maintains a small and nearly constant distance to the preceding vehicle~\cite{jia2016survey,cheng2015routing,li2014survey}. 
In particular, Land Transport Authority in Singapore has planned to build dedicated smart highway lanes, on which wireless connected vehicles move in platoons to increase roads' throughput~\cite{singaporfuturghighway}. The US Department of Transportation has developed ``The Automated Highway System'' so that vehicles can be driven in a platoon-like tight formation~\cite{usafuturehighway}. 

\begin{figure}[htb]
\centering
\includegraphics[width=4.5in]{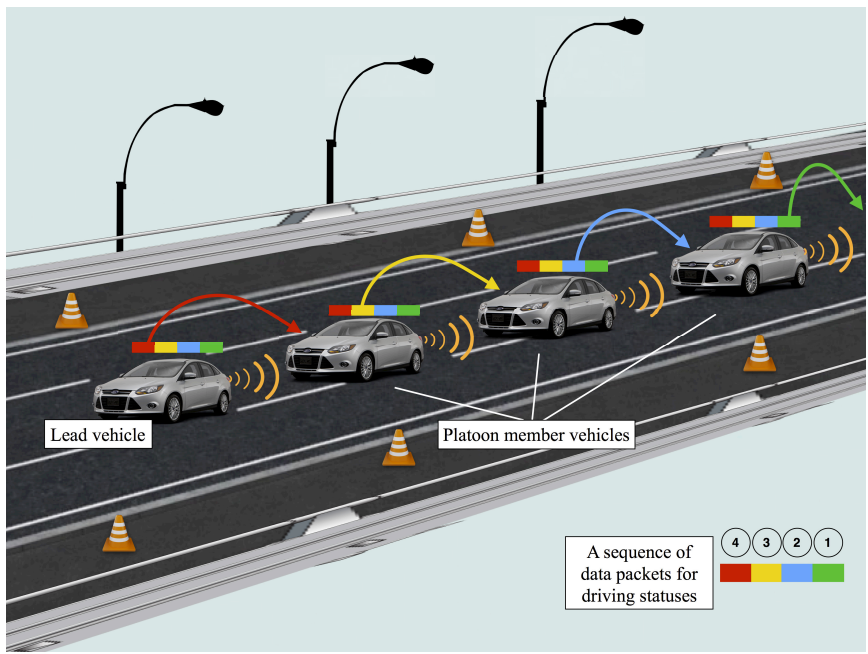}
\caption{A platoon of vehicles, where the driving status information is disseminated from the lead vehicle to the following automatic vehicles.}
\label{fig_app}
\end{figure}

Forming a vehicular platoon is shown in Figure~\ref{fig_app}. The lead vehicle decides the platoon's driving status, i.e., driving speed, heading directions, and acceleration/deceleration values, which depends on emergent road conditions, such as traffic jams, crossroads, obstacles or car accidents~\cite{wang2015performance}. The lead vehicle (managing the platoon) periodically broadcasts driving commands carrying information on its vehicle position and velocity to update the other platoon's vehicles. 
Specifically, the following vehicles act as command-forwarding nodes in a multi-hop vehicular network so that driving messages from the leader can be disseminated to all vehicles in the platoon. In terms of inter-vehicle communication, it is also assumed that the preceding vehicle disseminates driving commands to its following vehicle based on short-range one-hop broadcasts without causing interference to the other vehicles throughout the platoon~\cite{hartenstein2008tutorial}.

Inevitably, pushing vehicles to drive in close formation as the platoon requires low latency driving command transmission from the lead vehicle to the tail for driving safety. Two critical challenges arise in the inter-vehicle wireless communication. 
The first challenge is that signal fading induces dynamic wireless channels, which causes command loss at the receiver. This command loss is especially crucial in vehicular platoons since command reception at each vehicle highly depends on the reception of its preceding vehicle. Moreover, command loss at preceding vehicles can impact the command dissemination due to retransmissions. This may lead to fatal accidents to the rest of the platoon due to lack of timely updates. 
The second challenge is the possibility of assigning the exact transmit rate to each vehicle in the platoon. Although a high transmit rate achieves low transmission latency for each vehicle, increasing the transmit rate results in increasing the receiver's bit error rate (BER) at a given Signal-to-Noise Ratio (SNR). Accordingly, the vehicle with high BER spends longer time on command retransmissions, which prolongs dissemination latency of the platoons. Therefore, allocating the transmit rate without a proper adaptivity leads to command dissemination latency performance degradation. 

In this paper, we propose a low-latency driving command dissemination (LCD) algorithm to adapt the transmit rate (i.e., modulation) allocation of vehicles as such that the latency of command dissemination in the platoon is minimized under guaranteed BER. We prove that LCD algorithm achieves computation time complexity of $O(NM^2)$, where $N$ and $M$ are the number of vehicles and modulation levels, respectively. 
To quantify the command dissemination latency over time-varying channels, we interpret the vehicular platoon as a M/M/1 queue, where the length of the part of the platoon that have successfully received the driving command is modeled to be the length of the queue. 
In particular, we consider that the platoon is experiencing the identical channel fading at a specific time. The closed-form expected latency of the command dissemination is derived given a fixed command arrival rate and service rate. 

The rest of the paper is organized as follows. 
Section~\ref{sec_relatedwork} presents the related work on data transmission techniques in vehicular networks. 
Section~\ref{sec_system} introduces the system model of the vehicular platoon. 
In Section~\ref{sec_algorithm}, we propose the LCD algorithm based on dynamic programming technique. Moreover, the command dissemination latency is analyzed by using the Markov chain model with the M/M/1 queue. 
Simulation results are shown in Section~\ref{sec_evaluation}, followed by the conclusion in Section~\ref{sec_cond}.

%=============================================================================%
%============================Section 2 Related Work===============================%
\section{Related Work}
\label{sec_relatedwork}
In~\cite{togou2016throughput}, two Markov chains are used to model IEEE 802.11p EDCA throughput over service channels. An accurate model of the normalized throughput is obtained for each access category. 
%In~\cite{saleet2011intersection}, a geography-based communication protocol is studied to select road intersections through which a packet must pass to reach a gateway to the Internet. Geographical forwarding is used to transfer packets between any two intersections on the path, reducing the path's sensitivity to individual vehicle movements. 
The extensive datasets of vehicular traces are utilized by the conditional entropy analysis to unveil that there exists strong spatiotemporal regularity with vehicle mobility~\cite{zhu2014trajectory}. By extracting mobility patterns from historical vehicular traces, vehicular trajectory predictions are used to derive the packet delivery probability. 
Unfortunately, these communication protocols route data to the connected vehicles in terms of the availability of global geographical information, i.e., the position of vehicle and the driving map, which can not be applied to the vehicular platoon with only local channel information. 

Several communication protocols~\cite{khabbaz2013delay,li2016energy,al2012new} are developed for data dissemination in vehicular networks without global information. 
A delay-aware data forwarding scheme in~\cite{khabbaz2013delay} provides a bundle delivery delay in the context of a two-hop vehicular intermittently connected network. Moreover, the scheme allows a source roadside unit to carry out bundle retransmissions to high speed vehicles newly entering its communication range. In turn, these vehicles guarantee delay-minimal delivery of the retransmitted bundles to the data destination. 
In~\cite{li2016energy}, an energy-efficient cooperative relaying scheme is studied to extend mobile network lifetime while guaranteeing the success rate. The optimal transmission schedule is formulated to minimize the maximum energy consumption under guaranteed bit error rates. Furthermore, a computationally efficient suboptimal algorithm to reduce the scheduling complexity, where energy balancing and rate adaptation are decoupled and carried out in a recursive alternating manner. 
A hybrid location-based data dissemination protocol that combines features of reactive routing with geographic routing is designed to address link failures caused by vehicles mobility~\cite{al2012new}. Such link failures require direct response from the dissemination protocol, leading to a potentially excessive increase in the communication overhead and degradation in network scalability. 

Different from the data dissemination approaches in literature, we focus on the adaptive transmit rate in the vehicular platoon while guaranteeing the successful command delivery.

%=============================================================================%
%============================Section 3 System Model===============================%
\section{System Model}
\label{sec_system}
Assume that there are $N$ vehicles of interest in the platoon, forming $(N-1)$ wireless hops. 
In particular, we consider that the platoon has already been formed and is traveling on a straight single-lane highway with no need to change the platoon size or perform maneuvers (split, merge, leave, etc). 
The driving commands that contain driving status information are generated at the lead vehicle. The commands are immediately forwarded to its following vehicle all the way to the tail vehicle of the platoon, using vehicle-to-vehicle communication. 
The transmit rate that vehicle $i$ ($i \in [0,N-1]$) uses to forward packets is $r_i$, which specifies the number of bits per symbol, and $r_i \in \{\rho_1, \rho_2, ..., \rho_M\}$; where $\rho_1$, $\rho_2$, and $\rho_3$ indicate binary phase-shift keying (BPSK), quadrature-phase shift keying (QPSK), and 8 phase-shift keying (8PSK), respectively, and $r_i \geq 4$ corresponds to $2^{r_i}$ quadrature amplitude modulation (QAM). $\rho_M$ is the highest-order modulation. 
Suppose that the BER requirement of vehicle $i$ is $\epsilon_i$. 
The required transmit power at vehicle $i$ that guarantees the BER of $\epsilon_i$ is given by~\cite{he2014optimal}
\begin{equation}
P_i \approx \frac{\kappa_2^{-1}\ln\frac{\kappa_1}{\epsilon_i}}{\|\mathbf{h}_{i,i+1}\|^2}(2^{r_i}-1)
\label{eq_Txpower}
\end{equation}
where $\|\mathbf{h}_{i,i+1}\|$ denotes the channel amplitude between vehicles, $i$ and $(i+1)$ in the platoon. Without loss of generality, the inter-vehicular communication is typically based on the Rayleigh fading channel model~\cite{seyfi2011relay}. $\kappa_1$ and $\kappa_2$ are channel-dependent constants. 
Moreover, $\gamma_{i,i+1}$ defines SNR of the channel between vehicles $i$ and $(i+1)$, which is given by   
\begin{align}
\gamma_{i,i+1} = \frac{\|\mathbf{h}_{i,i+1}\|^2P_i}{\sigma^{2}_{0}}
\end{align}
where $\sigma_{0}$ is the channel noise power. 

Consider independent Rayleigh fading channel between the two vehicles in the platoon. Successful packet transmission probability at vehicle $i$, which indicates a successful packet reception at vehicle $(i+1)$, can be given by 
\begin{align}
\eta_{i} &= \mathbb{P}(\gamma_{i,i+1} \geq \bar{\gamma}_{i})
\label{eq_tranprob}
\end{align}
where $\bar{\gamma}_{i} = 2^{r_i} - 1$. 

In terms of medium access, an error-control stop-and-wait protocol is applied to the inter-vehicle communication. Specifically, vehicle $(i+1)$ replies an acknowledge (ACK) message to its preceding vehicle $i$ once a driving command is successfully received. A non-ACK message is replied by vehicle $(i+1)$ if the transmission is not successful, which leads to packet retransmissions at vehicle $i$. 
Note that the driving status information is critical, which needs to be forwarded hop-by-hop to ensure driving safety and controllability of every vehicle in the platoon. Therefore, we consider that the command is always routed to the nearest following vehicle as the next hop.

%=============================================================================%
%============================Section 4 Command Dissemination Problem====================%
\section{Command Dissemination Algorithm and Latency Analysis}
\label{sec_algorithm}
In this section, LCD algorithm is studied to determine transmit rate $r_i$ for each vehicle $i$ in the platoon according to dissemination latency and channel fading, which achieves the minimum driving command dissemination latency from the lead vehicle to the tail. Furthermore, we analyze the latency of dissemination in the vehicular platoon by using M/M/1 queuing theory~\cite{stewart2009probability}. 

\subsection{LCD Algorithm for Vehicular Platoon}
The structure of an optimal solution can be characterized based on a tradeoff, where an excessively high modulation could lead to high BER and large power consumption, but small transmission latency, while an excessively low modulation would result in a long transmission delay but small packet loss and power consumption. 
Here, our key idea is to allocate the optimal transmit rate from the lead vehicle to the $(N-1)$th vehicle (namely, $\boldsymbol{r^{\star}} = [r_0^{\star},r_1^{\star},r_2^{\star} ...,r_{N-1}^{\star}]$) to minimize the command dissemination latency. 
We present LCD algorithm based on dynamic programming approach to produce an optimal joint control policy for the transmit rate allocation in the platoon. 

Let $T_i$ denote the time that vehicle $i$ spends on command dissemination. The dissemination latency on each vehicle in the platoon can be given by 
\begin{align}
T_i = T_{i-1} + \frac{L}{r_i}
\end{align}
where $L$ is the packet length. Especially, the transmission time of the lead vehicle gives $T_0 = \frac{L}{r_0}$. 
Due to lossy channels, taking actions $r_i \in [\rho_1,\rho_M]$ at vehicle $i$ leads to two possible outcomes: 1) the command is successfully sent to the following vehicle, and the transmission time at $i$ is $T_i = \{T_i + T_{i-1} | r_{i-1}\}$; 2) the command is not successfully transmitted, and $T_i = T_{i-1}$. 
Then, we recursively define the subproblem for vehicle $i$, which is given by
\begin{align}
T_i(r_i) = \min\Big\{ \sum^{i}_{j=1} T_j(r_j) \Big| T_{j-1}(r_{j-1}),& \sum^{i}_{j=0}\frac{P_j}{i} \leq P_{max} \Big\}, \nonumber \\
&\forall i \in [1,N]
\label{eq_minTi}
\end{align}
$T_i(r_i)$ can be solved recursively based on the results of preceding subproblems at vehicle $(i-1)$ according to Bellman equation. 
Moreover, the optimal transmit rate $r_i^{\star}$ for minimizing the command dissemination latency is derived by all solutions to the preceding subproblems of $v_{i-1}$, which is  
\begin{align}
r_i^{\star} = \text{argmin}\{\sum^{N-1}_{i=0} T_i(r_i) | \sum^{N-1}_{i=0}\frac{P_i(r_i)}{N} \leq P_{max}\}
\label{eq_optimalRate}
\end{align}

The proposed LCD is presented in Algorithm~\ref{alg_LCD}, which derives the optimal actions, i.e., $r_i^{\star}$, by conducting backward induction in dynamic programming~\cite{cormen2009introduction}. Specifically, LCD iterates over the action, and uncertainty spaces for each action stage to calculate the exact latency function and corresponding policy function in each action stage. 
Moreover, LCD solves each subproblem of~\eqref{eq_minTi} just once and then saves its solution in a table, thereby avoiding recomputing the solution every time it solves each subproblem. 
\begin{algorithm}[t]
\begin{algorithmic}[1]
\caption{LCD Algorithm}
\label{alg_LCD}
\STATE{\textbf{Initialize:} $P_i$, $r_i = \rho_M$, $T_i(0) = 0$, $T_i(r_i) = \infty$, and $T_{total} = 0$. Platoon length = $N$.}
\STATEx{\textbf{Dynamic Programming}} 
\STATE{Initialize: $T_i(r_i) = \infty$ and $T_i(0) = 0$.}
\FOR{$i \to N$}
\FOR{$r_i \to [\rho_1, \rho_M]$}
\STATE{$T_i(r_i) \gets \min\Big\{ \sum^{i}_{j=1} T_j(r_j) \Big| T_{j-1}(r_{j-1}),$}
\STATEx{~~~~~~$\sum^{i}_{j=1}\frac{P_j}{i} \leq~P_{max} \Big\}$}
\STATE{Save the results of $T_i(r_i)$ while updating $\boldsymbol{r^{\star}}$.}
\ENDFOR
\ENDFOR
\STATE{The optimal solution is given by }
\STATEx{~~$r_i^{\star} \gets \text{argmin}\{\sum^{N}_{i=0} T_i(r_i) | \sum^{N}_{i=0}\frac{P_i(r_i)}{N} \leq P_{max}\}$.}
\STATEx{\textbf{Backward induction}}
\STATE{Initialize: $r_i = r_i^{\star}$.}
\FOR{$i = N \to 1$}
\STATE{Power control: $\sum^{i}_{j=1}\frac{P^\star_j}{i} \leq P_{max}$.}
\STATE{Trace backward: $T_{total} = T_{total} + T_i(r_i^{\star})$}
\ENDFOR
\STATEx{\textbf{Power control}}
\STATE{The transmit power of each vehicle, $P^\star_i$, is obtained by Equation~\eqref{eq_Txpower} with $r_i^{\star}$.}
\end{algorithmic}
\end{algorithm}

Turning now to the complexity of Algorithm~\ref{alg_LCD}. LCD iteratively searches the optimal solution in~\eqref{eq_optimalRate} to a given instance of the problem in~\eqref{eq_minTi} by using the backward induction in dynamic programming. LCD proceeds by finding the optimal solution $r_i^{\star}$ for each subset of attributes in a bottom-up fashion, utilizing the principle of optimality to reduce unnecessary computation. Specifically, going from vehicle $i+1$ backward to $i$ requires $M^2$ elementary computations. Hence, the cost of completing the transmit rate allocation in LCD is $O(NM^2)$. 

\subsection{Platooning Dissemination Latency Analysis}
\label{sec_queue} 
Since each vehicle in the platoon can be regarded as a server with infinite buffer, the command dissemination from the lead vehicle to the tail vehicle can be formulated with M/M/1 queuing discipline. 
In particular, suppose the command arrival rate at vehicle $i$ is $\lambda_i$, and the processing rate at the vehicle $i$ is $\mu_i$. Driving commands are examined on a first-come-first-served basis. 
When dissemination states, i.e., command arriving and processing, at all vehicles are steady, we have the following transition matrix, 
\begin{align}
\textbf{T}=&\left[
\begin{array}{l}
\lambda_0~~~~~~~-\mu_1~~~~~~~~0~~~~~~~~0~~~~~~~0~~\cdot \cdot \cdot~~~0~~0\\
-\lambda_0~~(\mu_1 + \lambda_1)~~~~-\mu_2~~~~~0~~~~~~~0~~ \cdot \cdot \cdot ~~~0~~0\\
0 \;\;\;\;\;\;\;\;\;\;\;\; -\lambda_1 \;\;\; (\mu_2 + \lambda_2) \;\; -\mu_3 \;\;\;\; 0 \;\;\;  \cdot \cdot \cdot \;\;\; 0 ~~ 0\\
\cdot \;\;\;\;\;\;\;\;\;\;\;\; 0 \;\;\;\;\;\;\;\;\;\;\;\;\;\;\; -\lambda_2 \;\;\;\;\;\;\; \cdot  \\
\cdot \;\;\;\;\;\;\;\;\;\;\;\; \cdot \;\;\;\;\;\;\;\;\;\;\;\;\;\;\;\;\; 0 \;\;\;\; \;\;\;\;\;\;\; \; \cdot \\
\cdot \;\;\;\;\;\;\;\;\;\;\;\; \cdot \;\;\;\;\;\;\;\;\;\;\;\;\;\;\;\;\; \cdot \;\;\;\; \;\;\;\;\;\;\; \;\;\;\;\;\; \cdot  \\
\cdot \;\;\;\;\;\;\;\;\;\;\;\; \cdot \;\;\;\;\;\;\;\;\;\;\;\;\;\;\;\;\; \cdot \;\;\;\; \;\;\;\;\;\;\;\;\;\;\;\;\;\; \;\;\;\; \cdot \\
\end{array}
\right]
\label{eq_T}
\end{align}

Let $\Pi_{i}$ denote the probability of the command being disseminated to vehicle $i$. Based on $\textbf{T}$ in~\eqref{eq_T}, the steady states of vehicles in the platoon are given by 
\begin{align}
\Pi_{i} = \lim_{t \to \infty} \Pi_i (t)
\end{align}
Resulting from the steady Markov chain with the transition matrix $\textbf{T}$, we have the following balance equation for a steady-state platoon, 
\begin{align}
-(1 - \Pi_{i})\mu_i + \Pi_{i-1}\lambda_{i-1} &= 0,~\forall i
\label{eq_balanceEq}
\end{align}
where $i = [1,+\infty)$ considering that there are $N$ vehicles in the platoon. 

In particular, the transmit rate allocated for each vehicle in the platoon is not correlated to each other due to the independent Rayleigh fading channel that is considered in the inter-vehicle communication (Section~\ref{sec_system}). 
Therefore, in terms of the latency analysis of command dissemination, we consistently consider a generic case that the vehicles experience independent fading channel with a command arrival rate, $\lambda_i$, and processing rate, $\mu_i$. 
Moreover, we also consider that the vehicles experience identical fading channel, i.e., $\eta_i$ has an identical probability distribution. As a result, we have $\lambda_i = \lambda_c$ and $\mu_i = \mu_c$, where $\lambda_c$ and $\mu_c$ are the fixed command arrival rate and processing rate, respectively. 

Define $\phi$ as the utilization of the vehicle for the driving command, which is  
\begin{align}
\phi = \frac{\lambda_c}{\mu_c}
\label{eq_phi}
\end{align}
where $\lambda_0 = \lambda_1 = \lambda_2 = ... = \lambda_c$ and $\mu_0 = \mu_1 = \mu_2 = ... = \mu_N = \mu_c$. 
Moreover, based on queuing theory, we know that  
\begin{align}
\sum_{\forall i \in [0,+\infty]} \Pi_i = 1
\label{eq_Pisum}
\end{align} 
By substituting~\eqref{eq_balanceEq} and~\eqref{eq_phi} to~\eqref{eq_Pisum}, we have 
\begin{align}
&\frac{1 - (-\phi)^N}{1 + \phi} \Pi_0 = \phi - 1 - \phi^2 + \phi - 1 + ... +  \nonumber \\
&~~~~(-1)^{N} \phi^{N-1} + (-1)^{N-1} \phi^{N-2} + ... + \phi - 1
\label{eq_pi0equ}
\end{align}
It can be observed that the right hand side of~\eqref{eq_pi0equ} is an Arithmetico-geometric sequence. Therefore,~\eqref{eq_pi0equ} can be written as 
\begin{align}
\Pi_0 = \frac{N+\phi(N+1) - (-\phi)^N + (-\phi)^{N+1} + (-\phi)^N}{(1+\phi)(1 - (-\phi)^N)}
\end{align}

Furthermore, the expected number of commands in the platoon when the vehicles are in steady state can be given by 
\begin{align}
\mathbb{E} (\omega^{\prime}) = \sum^{N-1}_{i=0} i \mathcal{P}_{\omega^{\prime}} 
= \frac{\phi}{1-\phi}
\label{eq_identical_command}
\end{align}
where $\omega^{\prime}$ denotes the number of commands, and $\mathcal{P}_{\omega^{\prime}} = \phi^i (1 - \phi)$. 
The variation of expected number of commands is 
\begin{align}
Var(\omega^{\prime}) = \frac{\phi}{(1-\phi)^2}
\end{align}

Using \textit{Little's Law}, the average latency of command dissemination can be given by 
\begin{align}
\mathbb{E} (T^{\prime}) &= \frac{\mathbb{E} (\omega^{\prime})}{\lambda_c} = \frac{1}{\mu_c - \lambda_c}
\label{eq_Tmm1}
\end{align}

%=============================================================================%
%============================Section 6 Evaluation=================================%
\section{Numerical Evaluation}
\label{sec_evaluation}
In this section, we implement LCD in simulations to evaluate its performance, namely, command dissemination rate and latency. Here, the command dissemination rate defines the ratio of successfully disseminated commands to the time consumption on the dissemination. For comparison purposes, we simulate three other scheduling algorithms that are suitable in our context setting. 

\subsection{Simulation Settings}
The platoon size increases from 5 to 50, i.e., $N \in [5,50]$. We normalize the noise power at the vehicles as $\sigma^2_0 = 1$, set the average channel power gain of the Rayleigh fading inter-vehicle communication links to $\nu_i = 1$. Furthermore, set the target $\epsilon = 0.05\%$ for the numerical results, i.e., the number of bit errors is no more than 0.05\%, however, this value can be configured depending on the traffic type and data quality-of-service requirements. For  $P_i$ in~\eqref{eq_Txpower}, the two constants, $\kappa_1 = 0.2$ and $\kappa_2 = 3$. 
The highest modulation scheme $\rho_M = 8$. Each command that is generated by the lead vehicle has a payload of 32 bytes, i.e., $L = 256$, unless otherwise specified. 
Block fading is assumed on all the wireless links. In other words, the channel gain of a wireless link keeps constant during the rate allocation and the command transmission, but varies between time frames. This assumption is reasonable, because the duration of a frame is typically up to 10$ms$ during which the distance that a vehicle has moved in highway speed is negligible.

We compared LCD with two dissemination algorithms with a fixed transmit rate, i.e., the lowest transmit rate protocol (LTRP), and power-constrained transmit rate protocol (PCTRP). LTRP is a simple command dissemination algorithm, which sets the lowest modulation scheme $\rho_0$ to the platooning vehicles. PCTRP sets the transmit rate of the vehicles to the highest transmit rate that is restricted by the highest transmit power $P_{max}$. 
In addition, a partially adaptive protocol adopting an \textit{ON--OFF} transmit rate selection (NFTRP) is also simulated for comparison (extended from~\cite{fu2006optimal}). Specifically, the vehicle $i$ exclusively determines $r_i$ according to $\eta_{i,i+1}$, where vehicle $i \in [0,N-1]$ with a high $\eta_{i,i+1}$ selects the maximum transmit rate, $\rho_M$. Otherwise, the transmit rate of the vehicle $i$ is $\rho_1$. 

\subsection{Simulation Results}
We compare the dissemination rate and latency in terms of the platoon size. In particular, the dissemination rate defines the ratio of the amount of commands successfully disseminated by all the vehicles in the platoon to the time spent on the dissemination. The dissemination latency defines time duration for an head-to-tail delivery of a command. 

\begin{figure}[htb]
\centering
\includegraphics[width=5in]{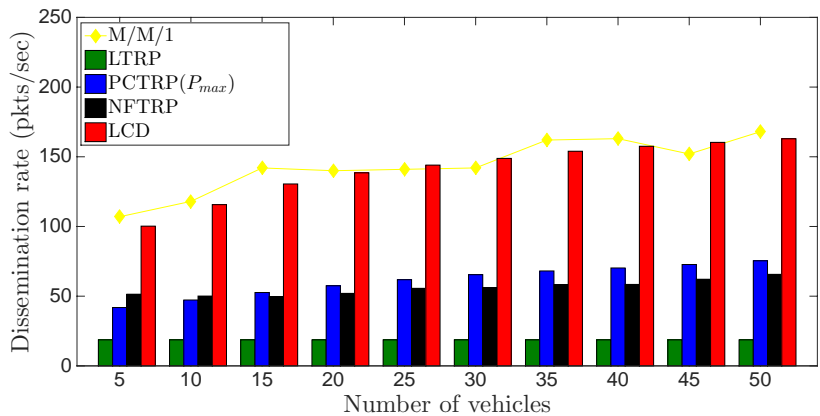}
\caption{The performance of command dissemination rate with different transmit rate allocation algorithms.}
\label{fig_car_rate}
\end{figure}

Figure~\ref{fig_car_rate} shows the dissemination rate of LTRP, PCTRP, NFTRP, the proposed LCD, and M/M/1 formulation in~\eqref{eq_Tmm1} with an increasing number of vehicles. Generally, the performance of LCD and PCTRP increases linearly with the platoon's size, while the one in LTRP does not vary too much. NFTRP provides a similar dissemination rate to PCTRP, which is better than LTRP. However, NFTRP with the \textit{ON--OFF} transmit rate selection is unable to efficiently adapt to the time-varying channel due to a hard link-quality threshold. 
In addition, LCD achieves 50.9\%, 87.6\%, and 64\% in terms of the PDCRP, LTRP, and NFTRP, respectively. 
In Figure~\ref{fig_car_rate}, it can also be observed that LCD approximates the numerical calculation of M/M/1 analysis. Therefore, the performance of M/M/1 also confirms validity of LCD. 

\begin{figure}[htb]
\centering
\includegraphics[width=5in]{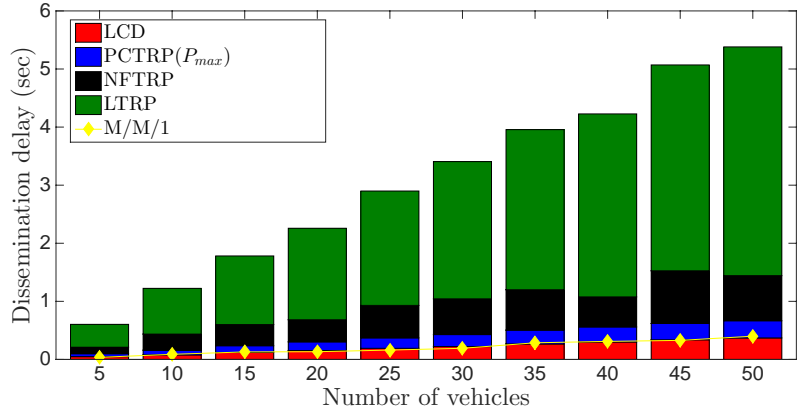}
\caption{A comparison of dissemination latency by our LCD, and the typical rate allocation strategies.}
\label{fig_car_delay}
\end{figure}

Figure~\ref{fig_car_delay} presents the average latency of disseminating commands from the lead vehicle. In particular, PCTRP gives the smallest dissemination latency since the highest transmit rate at all the vehicles leads to the lowest dissemination latency. However, note that a high transmit rate causes a high bit error rate, which reduces the command dissemination rate, as shown in Figure~\ref{fig_car_rate}. LCD approaches PCTRP and M/M/1 analysis with the maximum gap of up to 0.2 second in terms of latency, while LCD improves the command dissemination rate by 50.9\%, which is significant to guarantee the timeliness and reliability of driving information delivery in the platoon. 
In addition, LCD outperforms LTRP with substantial gains about 4.1 seconds, and the gains keep growing with the platoon size. LCD also achieves 1.2 seconds faster than NFTRP in terms of dissemination latency. 

\begin{figure}[htb]
\centering
\includegraphics[width=5in]{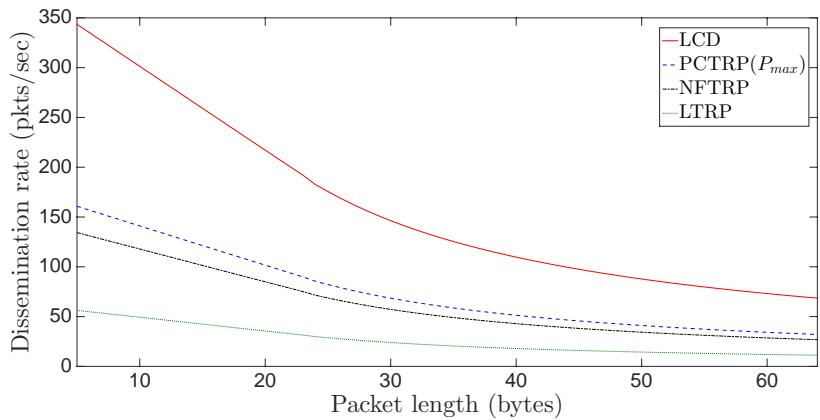}
\caption{A comparison of command dissemination rate using different rate allocation strategies.}
\label{fig_data_rate}
\end{figure}

\begin{figure}[htb]
\centering
\includegraphics[width=5in]{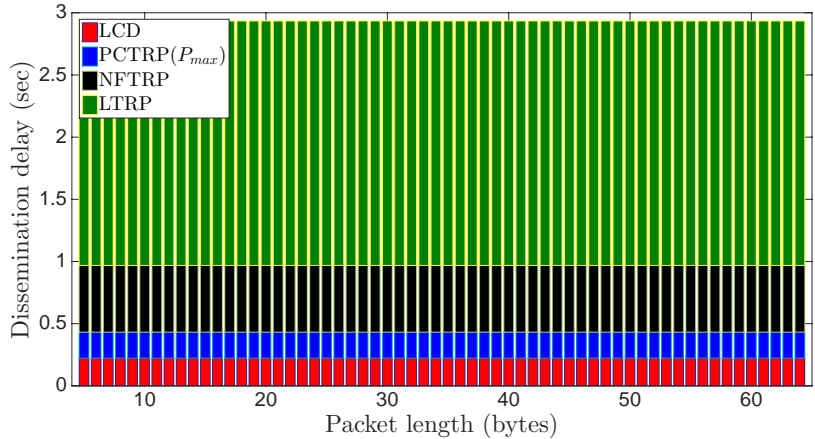}
\caption{A comparison of dissemination latency using LCD, and the typical rate allocation strategies.}
\label{fig_data_delay}
\end{figure}

Furthermore, we compare the dissemination rate, and latency in terms of command packet size, given a platoon of 30 vehicles. Figure~\ref{fig_data_rate} shows the performance of the command dissemination rate, where the command packet size increases from 5 bytes to 65 bytes. 
In general, the maximum dissemination rate of LCD is higher than PCTRP, LTRP, and NFTRP by around 56.3\%, 85.6\%, and 60\%, respectively. Additionally, the command dissemination rate of LCD drops with the increase of the packet size. The reasons can be explained by the fact that the number of disseminated commands drops with an increase of command packet size since each vehicle spends longer time on transmission. Moreover, in terms of the dissemination latency shown in Figure~\ref{fig_data_delay}, all the transmit rate allocation algorithms maintain a fixed dissemination latency regardless of the packet size. The reason is that a fixed number of vehicles in the platoon results in a lower bounded $T_i(r_i)$ in~\eqref{eq_minTi}, which is unaffected by the packet size. In particular, LCD saves 0.1, 2.7, and 0.8 seconds dissemination time compared to PCTRP, LTRP, and NFTRP. 

%=============================================================================%
%============================Section 7 Conclusion=================================%
\section{Conclusion}
\label{sec_cond}
In this paper, we address the optimal transmit rate allocation problem for driving command dissemination in vehicular platoons. A low-latency command dissemination scheme, LCD, is studied to apply dynamic programming to produce an optimal rate allocation policy. To further quantify the latency, we formulate the command dissemination in the platoon by using a M/M/1 queue model. 
Simulation results show that LCD significantly improves the dissemination rate by 50.9\%, as compared to the existing algorithms. Moreover, LCD also approximates the lower bound of dissemination latency with the maximum gap of up to 0.2 second.

\section*{Acknowledgement}
This work was partially supported by National Funds through FCT (Portuguese Foundation for Science and Technology) within the CISTER Research Unit (CEC/04234); also by FCT and the EU ECSEL JU under the H2020 Framework Programme, within project ECSEL/0002/2015, JU grant nr. 692529-2 (SAFECOP).

\ifCLASSOPTIONcaptionsoff
  \newpage
\fi

\bibliographystyle{IEEEtran}
\bibliography{LCD}

% that's all folks
\end{document}